# Optical-limiting-based materials of mono-functional, multi-functional and supramolecular $C_{60}$-containing polymers

Hendry Izaac Elim,[a*] Jianying Ouyang,[b] Suat Hong Goh,[b] and Wei Ji[a♣]

[a] Department of Physics, National University of Singapore,

3 Science Drive 3, Singapore 117543, Singapore

[b] Department of Chemistry, National University of Singapore,

3 Science Drive 3, Singapore 117543, Singapore

*Abstract*

By using nanosecond laser pulses at 532-nm wavelength, we have intensively studied the optical limiting and nonlinear optical properties of some new composites which consist of mono-functional 1,2-dihydro-1,2-methanofullerene[60]-61-carboxylic acid (FCA), multifunctional fullerenol, and supramolecular $C_{60}(HNC_4H_8NCH_3)_9$ - containing polymers. The optical limiting performance of FCA/poly(styrene-co-4-vinylpyridine) is found to be better than that of its parent $C_{60}$, while the multi-functional Fullerenol and supramolecular $C_{60}(HNC_4H_8NCH_3)_9$ incorporated with polymers show poorer optical limiting responses. The possible sources for the improvement in the optical limiting are discussed.

[*] Corresponding author, email address: phyehi@nus.edu.sg

[♣] or phyjiwei@nus.edu.sg

## 1. Introduction

Fullerene and its derivatives have attracted lots of interest for their unique interaction with polymers [1-7] and their wide range in promising applications as an excellent optical limiter for eye and sensor protection from high intensity laser pulses [8,9]. Other potential such as optical switches and a single-$C_{60}$ transistor has been investigated [10,11]. Moreover, recently strong magnetic signals have been found in rhombohedral $C_{60}$ polymer (Rh-$C_{60}$), and studies of synthetic route to the $C_{60}H_{30}$ polycyclic aromatic hydrocarbon and its laser-induced conversion into fullerene-$C_{60}$ have also received attention [12,13]. Hence, the preparation and both electronic and optical properties of fullerene composites are still interesting to be studied.

$C_{60}$-based polymeric materials are normally prepared by simply dispersing $C_{60}$ into polymeric matrices. However, the poor adhesion between $C_{60}$ and polymer prevents the well dispersion of $C_{60}$ in the matrix [14,15]. To improve the affinity therein, $C_{60}$ is functionalized and interacts with complementary functional groups of a polymer, and the specific interaction between $C_{60}$ and the polymer plays an important role in the resulting mechanical performance and other optical properties [14-16]. It is further expected that the optical properties of $C_{60}$ derivatives may change upon functionalization. So far, Sun and his colleagues [3,17,18] have examined a series of methano-$C_{60}$ derivatives and they have found that the mono-functional $C_{60}$ derivatives show similar optical limiting responses to that of parent $C_{60}$ while multi-functional ones give poorer performance. In addition, since the first example of $C_{60}$ involved in supramolecular phenomena was reported by Ermer in 1991 [19], many related researches [20-23] have been focused on the formation of supramolecular buckminsterfullerene, fullerene chemistry, and assemblies and arrays held together by weak intermolecular interactions. Moreover, the recent studies deal with supramolecular $C_{60}$-containing polymeric materials based on functionalized $C_{60}$ and polymers possessing suitable functional groups, which successfully overcome the incompatibility between pristine $C_{60}$ and polymers. However, the nonlinear optical (NLO) properties of similar supramolecular fullerene incorporated with polymers (PSI-46) have not been fully investigated yet. Consequently, the main challenge that still remains is how to improve the optical limiting and nonlinear optical properties of $C_{60}$.

Here, we report a systematic investigation of the excited state mechanisms correlated to nonlinear optical limiting properties of mono-functional 1,2-dihydro-1,2-methanofullerene[60]-61-carboxylic acid, multi-functional fullerenol and 1-(4-methyl)-piperazinylfullerene[60] –containing polymers in room-temperature solution measured under the same experimental conditions as those used for the fullerenes. The results show that the NLO responses of mono-functional FCA/PSVPy32 toward nanosecond laser pulses at 532 nm are better than that of the [60]fullerene, which suggests significant contributions of triplet-triplet states absorption and a fast intersystem crossing (ISC) associated with five-level mechanism of the fullerenes. While the mechanism of the multi-functional fullerenol, and MPF -containing polymers show that both the nonlinear optical and the optical limiting performances of the fullerenes are significantly affected by the disturbance of π–electronic system in $C_{60}$ cage due to multi-addends of fullerenol or MPF. Mechanistic implications of the experimental results are discussed, and an excited state reverse saturable absorption model that consistently accounts for the nonlinear optical properties of all the samples is explained.

## 2. Experimental Section

### 2.1. Materials

Our $C_{60}$ (99.9%) sample used in these researches was obtained from Beijing University, Beijing, China. While mono-functional 1,2-dihydro-1,2-methanofullerene[60]-61-carboxylic acid, multi-functional fullerenol and 1-(4-methyl)-piperazinylfullerene[60] –containing polymers were prepared in the following steps, respectively.

Firstly, 1,2-dihydro-1,2-methanofullerene [60]-61-carboxylic acid (FCA) was synthesized by the method reported by Isaacs and his co-workers [24,25]. The FCA has been prepared carefully to be [6,6]-closed isomer with 58 π-electrons. Poly (styrene-co-4-vinylpyridiene) (PSVPy) and polystyrene (PS) were prepared by free-radical copolymerization initiated by an initiator for polymerization (AIBN). The molar percentage of 4-vinylpyridine unit in PSVPy32 was 32% as determined by elemental analysis. $C_{60}$ and FCA were dissolved in toluene and 1,2-dichlorobenzene, respectively, into which appropriate amounts of PSVPy32 and PS were added. The six samples were prepared with the same transmittance of 65 % and concentration of ~$1.6 \times 10^{-3}$ M, as shown in Table 1. The structure of FCA [2] is shown below.

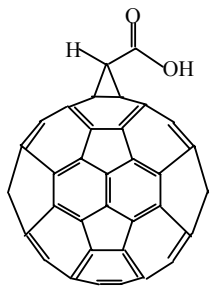

FCA

Secondly, multi-functional fullerenol (Fol) was synthesized and characterized according to literature [26]. On the average, Fol has 10 to 12 hydroxy addends [1]. Toluene and N,N-dimethylformamide (DMF) of AR grade were purchased from Fisher Scientific Company, USA and used as received. Two poly(styrene-co-4-vinylpyridine) (PSVPy) samples containing 20 and 32 mol% of vinylpyridine, denoted as PSVPy20 and PSVPy32, respectively, were prepared by free radical copolymerization. Poly(styrene-co-butadiene) was purchased from Aldrich Company, USA, which is a kind of high-impact polystyrene (HIPS) with a melt index (200$^{\circ}$C/5 kg, ASTM D 1238) of 2.8 g/10 min.

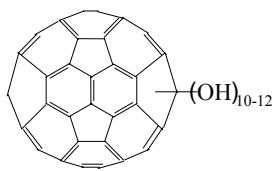

Fol

Fol was dissolved in DMF, into which an appropriate amount of PSVPy32 was added. The mixture was stirred continuously overnight and then added into diethyl ether dropwise with vigorous stirring. The precipitates (blends) were isolated by centrifugation and washed with diethyl ether for three times. The blends were dried in vacuo at 50 $^{\circ}$C for one week. The Fol contents in various blends were determined by thermogravimetric analysis. The brittleness of PSVPy made it difficult to prepare Fol/PSVPy samples for dynamic mechanical analysis (DMA) measurements [1]. Instead, a mixture of 26 parts of PSVPy20 and 74 parts of HIPS was used as the matrix. To achieve a better compatibility between PSVPy and HIPS, PSVPy20 was used in view of its higher styrene content. Appropriate amounts of Fol, PSVPy20 and HIPS were mixed in a Laboratory Mixing Molder (ATLAS, USA) at 190 $^{\circ}$C for 30 min at a speed of 120 rpm. The mixed samples were compressed into films with a thickness ca. 0.25 mm under a pressure of 12 MPa at 140 °C and then at room temperature at the same pressure for 30 min using a hydraulic press (Fred S. Carver Inc., USA). The Fol contents in various samples were determined by thermogravimetric analysis. Samples used in this study are PSVPy20/HIPS polymer matrix, and polymer matrix filled with 2.4, 4.8, and 11.6 wt% of Fol, which are denoted as Fol2.4, Fol4.8, Fol11.6, respectively.

Finally, supramolecular multi-functional 1-(4-Methyl)- piperazinylfullerene - containing polymers was prepared based on the following steps. 1-Methylpiperazine (98%) was purchased from Sigma-Aldrich Company, USA. (3-Cyanopropyl)methyldichlorosilane and dichlorodimethylsilane were supplied by Fluka Chemika-Biochemika Company. Chlorobenzene (AR grade) was purchased from BDH, UK, and tetrahydrofuran (THF, AR grade) from Fisher Scientific, UK. 1-(4-Methyl)-piperazinylfullerene (MPF), a red and powder-like multi-functional $C_{60}$ derivative, was synthesized and characterized according to literature [27], which has an average stoichiometry of $C_{60}(HNC_4H_8NCH_3)_9$ with a structure as follows:

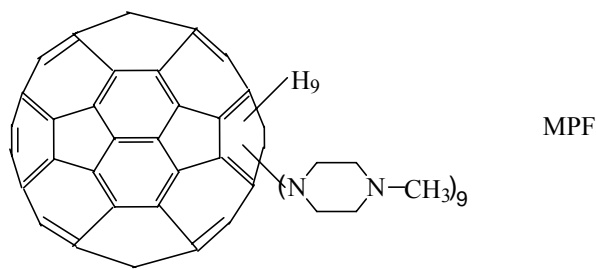

The random copolymer of dimethylsiloxane and (3-carboxypropyl)methylsiloxane was synthesized and characterized according to the method reported by Li and Goh [28]. It contains 45.5 mol% of (3-carboxypropyl)methylsiloxane unit as determined by $^1$H-NMR and has a number-average molecular weight 4,100 and polydispersity 1.08. This transparent and gel-like copolymer is denoted as PSI-46 with the following structure:

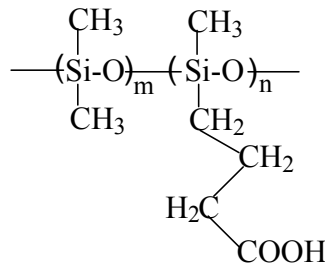

PSI-46

To prepare supramolecular multi-functional MPF/PSI-46 composites, an appropriate amount of PSI-46 was added into the THF solution of MPF. The mixture was continuously stirred overnight. Three samples were prepared, which are MPF/PSI-46 (1:2), MPF/PSI-46 (1:4), and MPF/PSI-46 (1:6), denoted as MPF(1:2), MPF(1:4), and MPF(1:6), respectively where the ratios in the parentheses refer to the ratios of nitrogen atoms of MPF over the carboxylic groups of PSI-46.

**2.2. Measurements**

The ultraviolet and visible (UV-vis) absorption sectra were recorded at the wavelength range ~190-820 nm on a Hewlett Packard 8452A Diode Array spectrophotometer with a Hewlett Packard Vectra QS/165 computer system. The optical limiting measurements were conducted using linearly polarized nanosecond optical pulses from a Q-switched, frequency-doubled Nd:YAG laser (Spectra Physics DCR3) with pulse duration of 7 ns or an optical parametric oscillator (Spectra Physics MOPO 710) with pulse duration of 5 ns. The spatial distribution of the pulses was nearly Gaussian after passing through a spatial filter. The pulse was divided by a beam splitter into two parts. The reflected part was taken as the reference representing the incident light energy and the transmitted beam was focused onto the sample by a focusing mirror (f = 25 cm). Both the incident and transmitted pulse energies were measured simultaneously by two energy detectors (Laser Precision RjP-735). The minimum beam waist of the focused laser beam was ~40 μm, determined by the standard Z-scan method [29]. The optical limiting (OL) measurements were conducted with the sample fixed at the focus. The OL measurements of FCA- and FOL-polymer composites were carried out with pulse duration of 7 ns. While the OL measurements of MPF-polymer composites were employed with pulse duration of 5 ns. In addition, the photoluminescence (PL) measurement for each sample was carried out using a luminescence spectrometer (LS 55, Perkin-Elmer Instrument U.K.) with the excitation wavelength of 442 nm for FCA-polymer composites, and 480 nm for multi-functional Fol and MPF – polymer composites, respectively.

## 3. Results and Discussion

### 3.1. Ultraviolet and visible (UV-vis) absorption spectra

UV-vis absorption spectra of the supramolecular mono- and multi-functional $C_{60}$-containing polymers were measured in room-temperature solutions. The spectral parameters are summarized in Table 1. For the mono-functional FCA-polymer composites, the absorption spectra are quite similar to that of $C_{60}$ as shown in Figure 1(a). The multi-functional Fol has a marginally different absorption spectrum in comparison with its parent $C_{60}$ in which the spectrum is slightly shifted to the blue shift. In contrast, the MPF has a significantly different spectrum from that of $C_{60}$. The difference is not only in spectral shape but also in absorptivity (Figure 1(c), and Table 1). This indication is caused by the broken symmetry of the π–electronic system in $C_{60}$ cage due to multi-addends of MPF.

### 3.2. Ground-State Absorption Parameters

The ground-state absorption parameters of the supramolecular mono-functional FCA, and multi-functional Fol and MPF - containing polymers at room temperature are shown in Table 1. The parameters are noticeably different from that of $C_{60}$. At 532 nm, the absorptivity ($\varepsilon$) of FCA and Fol -polymer composites is larger than that of its parent $C_{60}$. This contributes to the higher ground state cross section of the supramolecular in comparison with that of $C_{60}$. The similar observation has also been investigated by Sun et al. [3] in a series of fullerene derivatives. However, at the same wavelength the molar absorptivity of MPF observed at higher transmittance of 75% is much more poorer than that of its parent, indicating the effect of broken symmetry in the π–electronic system of $C_{60}$ cage. Moreover, the ground-state absorption spectra as displayed in Figure 1(c) of the MPF-polymer composites provide no evidence for molecular aggregation.

### 3.3. Photoluminescence performance of FCA, Fullerenol, and MPF – containing polymers versus $C_{60}$

Photoluminescence spectra of the FCA, Fol, and MPF - containing polymers were measured in room-temperature with different concentration of polymers. As shown in Figure 2(a), all mono-functional FCA incorporated with PSVPy or PS has poorer PL intensity than that of the multi-functional Fol and MPF – polymer composites. In general, one can also observe that most of the fullerene derivatives have an improvement of the PL by the reducing of the polymers concentration (Figures 2(b) and 2(c)). In addition, for the supramolecular multi-functional FoI and MPF, the peak of PL spectra are apparently shifted to the blue-shifted region in comparison to that of its parent $C_{60}$ (Figure 2). The results are consistent with that based on the UV-vis measurement in Fig. 1.

To evaluate qualitatively that there is light emission from the lowest excited singlet state to the ground state, that may affect the population of excited electrons at the excited triplet states, we have performed a series of photoluminescence (PL) measurements at room temperature. Figure 2(a) shows that there is an increase in the PL intensity of mono-functional FCA, in comparison to its parent $C_{60}$. This is due to a significant disturbance of π-electron system of $C_{60}$ cage upon derivatization, consistent with our optical limiting (Figure 3). Figure 2(a) also shows that the presence of polymer, either PSVPy32 or PS, in FCA slightly reduces the PL intensity. Moreover, the PL spectra show that there is polymer concentration dependence: the lower the polymer concentration is, the higher the PL intensity is. This implies an enhancement in the emission from the first excited singlet state to the ground state. Consequently, it decreases the probability of transferring electrons to the lowest excited triplet state by inter-system crossing (Figure 4), leading to poorer optical limiting behaviour. In addition, it should be noted that the PL emission of $C_{60}$ and FCA/polymer composites still exhibits very weak luminescence at room temperature due to high molecular symmetry of $C_{60}$. However, their PL may increase dramatically as they are cooled to low temperature because of reduction of thermal quenching of excited states [30].

In figures 2(b) and 2(c), the PL signals of multifunctional FoI and MPF –polymer composites are blue-shifted in comparison with their parent $C_{60}$. This is as the results of the broken symmetry of the π–electronic system in $C_{60}$ cage due to multi-addends of Fol or MPF. Furthermore, it may reduce most likely the triplet-triplet absorption during the laser radiation in the optical limiting measurements. According to the PL spectra, the more addends attached into the cage of $C_{60}$ are, the higher PL spectra are. As observed in monofunctional FCA, the PL spectrum of supramolecular multi-functional Fol – containing polymers was subsequently also concentration dependence. In contrast, it is interesting to notice that the PL signal of multifunctional MPF is marginally lower than that of MPF incorporated with polymer of PSI-46 which means that the additional polymer has enhanced the light emission from the lowest excited singlet state to the ground state of the MPF states. However, the PL signal was slightly diminished as the increase of the concentration of PSI-46. This process was similar to that of the mono-functional FCA.

## 3.4. Nonlinear optical limiting behaviors of FCA, Fullerenol, and MPF – containing polymers versus $C_{60}$.

For nanosecond laser pulses, transient reverse saturable absorption (RSA) induced by $C_{60}$ and its polymer composites can be described into five-level system [31-33] as displayed in Figure 4(a). Therefore, we can approximately formulate a sequential two-photon absorption (STPA), defined as

$$\alpha = \alpha_0[S_0] + \beta_{eff}I, \tag{1}$$

where

$$\beta_{eff} = \alpha_0\alpha_T[S_0], \tag{2}$$

$[S_0]$ is the ground state concentration, $\alpha_0$ and $\alpha_T$ are the absorption coefficients describing the ground-state ($S_0$) and triplet-state ($T_1$) absorption, respectively. Moreover, the triplet state concentration at instant $t$ can be approximated by $[T_1] = \alpha_0[S_0]I$. Such a process can be investigated by the Z-scan method [29]. The detailed NLO measurements of FCA, Fullerenol, and MPF – containing polymers can be found in references [2,4,34].

Optical limiting (OL) properties of the supramolecular mono- and multi-functional $C_{60}$ - containing polymers were investigated to compare the results with those of $C_{60}$. Shown in Figure 3 are optical limiting responses of the supramolecular in solutions of 65, 70 and 75% linear transmittances in a cuvette with 1 or 10 mm optical path length for concentrated or diluted samples. The transmittances are first linear with input fluences ($F_{in}$) and then go down as the nonlinear OL process called the RSA mechanisms happened at high input fluences.

To figure out a complete process of the unimolecular optical limiting performance happened in the supramolecular mono- and multi-functional $C_{60}$, we have adopted a five-level RSA model (Figure 4(a)), which in the absence of contributions from any bimolecular excited-state processes, as shown in Figures 4(b) and 4(c) to illustrate the excited state processes, specifically excited state absorption and emission including photoluminescence process. In the experiments of photoluminescence and RSA, the selected excitation corresponds to the $S_0 \rightarrow S_1$ transition, i.e. from the ground to the first excited state. The fullerene molecules are excited, with the selected excitation, to the first excited state of $S_1$ and its population can be probably accumulated. The $S_1 \rightarrow T_1$ intersystem crossing (ISC) populates the first excited triplet state $T_1$. Both the $S_1 \rightarrow S_2$ and $T_1 \rightarrow T_2$ transitions might cause absorption enhancement with elevated excitation (RSA effect). However, the transition probability from $S_1$ to $S_2$ can be neglected due to the use of nano-second pulses laser. Therefore, the upconverted emission attributed to the radiative $S_2 \rightarrow S_0$ transition can be ignored.

Having considered dymanical processes of RSA at the selected excitation of $S_0$, $S_1$, $T_1$ and $T_2$, the proposed dynamic equations for the supramolecular mono- and multi-functional fullerene are then expressed as follows

$$dn_{S0}/dt = -(\sigma_G I/h\nu)n_{S0} + n_{S1}/\tau_{SG} + n_{T1}/\tau_{TG}$$

$$dn_{S1}/dt = (\sigma_G I/h\nu)n_{S0} - (1/\tau_{SG} + 1/\tau_{ISC} + \sigma_{21} I/h\nu)n_{S1} + n_{S2}/\tau_{21} \quad (3)$$

$$dn_{T1}/dt = n_{S1}/\tau_{ISC} - (1/\tau_{TG} + \sigma_T I/h\nu)n_{T1} + n_{T2}/\tau_{TT}$$

$$dn_{T2}/dt = (\sigma_T I/h\nu)n_{T1} - n_{T2}/\tau_{TT}$$

where $n_i$ stands for population in the $i$th energy level, $I$ and $\nu$ are the excitation light intensity and frequency, $1/\tau_i$ or $1/\tau_{ij}$ is the decay rate from the $i$th level to the ground state or from the $i$th to the the $j$th level, $\sigma_G$, and $\sigma_{TT}$ are the absorption cross sections, corresponding to the $S_0 \rightarrow S_1$, and $T_1 \rightarrow T_2$ transitions. These populations should satisfy the relation (for $n_{S2}(t) \approx 0$)

$$n_0(0) = n_{S0}(t) + n_{S1}(t) + n_{S2}(t) + n_{T1}(t) + n_{T2}(t), \quad (4)$$

and the initial condition,

$$n_0(0) = n_{S0}(t), \quad (5a)$$

and

$$n_{S1}(0) = n_{S2}(0) = n_{T1}(0) = n_{T2}(0) = 0. \quad (5b)$$

These initial condition indicate that all molecules are in the ground state before excitation. Inducing nano-second optical pulses, propagating through the sample, can happen the excited state process due to a RSA. The propagation equation can be formulated as

$$\begin{aligned} dI/dz &= -(\alpha_0 S_0 + \beta_{eff} I)I \\ &= -(\alpha_0 S_0 + \alpha_T T_1)I \\ &\approx -(\sigma_G I/h\nu)n_{S0} - (\sigma_{21} I/h\nu)n_{S1} - (\sigma_{TT} I/h\nu)n_{T1} \end{aligned} \quad (6)$$

where $\beta_{eff}$ is defined in Eq.(2).

In general, the solution of Eq. (6) might be numerically solved using a computer when given specifications of the material parameters and the optical pulse features [29]. To simplify the equations based on the present experimental results one needs to assume a phenomenological description that the total absorption cross section $\sigma$ of the supramolecular mono- or multi-functional $C_{60}$ is a function of the laser fluence ($F$). Then the $\sigma(F)$ can be represented in the form of a power series of $F$ as follows [35]

$$\sigma(F) = \sigma_G + \mu_1 F + \mu_2 F^2 + \mu_3 F^3 + ... \quad (7)$$

Taking the first two terms into Beer's law and integrating it, one can obtain the equation of the transmittance input fluence-dependent

$$T = T_0 / [1 + (1 - T_0)(F_{in}/F_{NLO})], \quad (8)$$

where $T_0 = \exp(-\sigma_G n_0 L)$ is the sample transmittance in the limit of low light intensity, $L$ is the sample thickness, $F_{in}$ is the incoming laser fluence, and $F_{NLO} \cong \sigma_G/\mu_1$ is the parameter characterizing the nonlinear absorption of the material. In the case of high fluences, the ground state population is largely depleted, and the excited state population is distributed between the first excited-singlet and lowest-triplet states. A figure of merit for such RSA molecules involving both excited states can be defined as [36]

$$\sigma_{eff}/\sigma_G = \ln(T_{SA})/\ln(T_0) \quad (9)$$

where $T_{SA} = \exp(-\sigma_{eff} n_0 L)$ is defined as a saturated transmission for high degrees of excitation, and $\sigma_{eff}$ is an effective excited-state cross section.

To distinguish the different between saturable absorption and reverse saturable absorption in our experiment results, we can express the absorption coefficient in Eq. (1) as a function of excitation intensity

$$\alpha \cong \sigma_G /(1+I/I_{SA}) + \sigma_{eff}(I/I_{SA})/(1+I/I_{SA}). \tag{10}$$

The profile of $\alpha$ versus $I$ is remarkably dependent on the ratio of $\sigma_{eff}/\sigma_G$ as seen in our optical limiting data in Figure 3. The turning point at $\sigma_{eff}/\sigma_G = 1$ is apparently found where the system absorption keeps a constant regardless of excitation increase, and this indicates that there exists a balance of the absorption between the ground state and the excited states. The saturation absorption is found only if $\sigma_{eff}/\sigma_G < 1$, showing that the ground state absorption is stronger than the excited states at certain excitation wavelength. On the opposite side, the reverse saturable absorption appears while $\sigma_{eff}/\sigma_G > 1$, indicating that the absorption ability of the excited states is superior to that of the ground state.

In our observation, the optical limiting responses of the dilute $C_{60}$ solution are in general much weaker than those of the more concentrated $C_{60}$ solutions [1], similar to that observed for the supramolecular FCA, Fol and MPF –polymer composites. From figure 3, we have extracted optical limiting parameter such as limiting threshold of mono- and multi-functional fullerene as depicted into Table 2. The smaller limiting threshold is, the better optical limiting property is. Furthermore, the changes in optical limiting responses of the supramolecular multifunctional Fol and MPF –containing polymers with certain amount of polymer concentration are relatively small so that the polymer concentration dependence of optical limiting is not so obvious. In the present study, however, the high polymer concentration (PSVPy32) in mono-functional FCA has significantly improved the optical limiting performance of this mono-functional $C_{60}$. It is also unlikely that the results of PSVPy32 concentration dependence are due to molecular aggregation effects, because not only does the FCA-polymer composites have much better solubility characteristics than the parent $C_{60}$ but also some of the concentrations used in the measurements should be considered as fairly dilute.

Table 2 shows the nonlinear values of $F_{NLO}$ and $\sigma_{eff}/\sigma_G$ calculated based on the best experimental fitting of the data in figure 3. While the value of $T_{SA}$ was estimated from the OL data. For all solvents tested using nano-second pulses in this study, the nonlinear absorbing parameter $F_{NLO}$ of FCA/PS-c and FCA/PSVPy32-A is smaller than that of other samples. This result indicates that the nonlinear absorption becomes more efficient as the mono-functional FCA incorporated with a high concentration of polymer. The optical limiting behavior of the supramolecular multifunctional Fol and MPF – containing polymers may be similarly considered in the same mechanistic framework (Figure 3(a)). The intersystem crossing yield of the multi-functional is also unity, and there is also the ratio $\sigma_{eff}/\sigma_G > 1$ for the supramolecular, confirming optical limiting contributions from the unimolecular reverse saturable absorption mechanism shown in figure 3. More quantitatively, however, the ground-state absorption cross section of the Fol incorporated with polymers is larger than that of $C_{60}$ at 532 nm (Table 2) and the triplet-triplet absorption cross section of the supramolecular is smaller than that of $C_{60}$ at the same wavelength (Figure 3). Thus, $(\sigma_T/\sigma_G)_{multifunctional} < (\sigma_T/\sigma_G)_{C60}$, which would suggest weaker optical limiting responses for the supramolecular multi-functional of Fol or MPF –containing polymers in the context of the excited state mechanism. In addition to the high light intensities, the reduction of triplet-triplet absorption depends on how big the disturbance of π–conjugated electronic system in $C_{60}$ cage. The more the functional attached to the cage, the lower the triplet-triplet state absorption. This was confirmed in the relationship of the optical limiting results of their parent $C_{60}$ in room-temperature toluene or chlorobenzene (Figure 3) with the excited state model. Consequently, one can obtain that the limiting threshold of $C_{60}$ in chlorobenzene (2.7 J/cm$^2$) is better than that in toluene (3.1 J/cm$^2$). We suggest that this is due to nonlinear scattering contributed from the solution of chlorobenzene.

To examine the relative relationship between the lifetime of each concerned level in the supramolecular fullerene and the nano-second pulse width of the pumping laser, we assumed that the electronic relaxation of higher-lying excited states to the first excited state in the singlet or triplet manifold is very fast, typically on the scale of picoseconds or shorter [31]. Therefore, one can consider only the populations of the $S_0$, $S_1$ and $T_1$ manifolds due to the use of nano-second pulses in our experiments. Hence, the populations of the $S_2$ and $T_2$ manifolds are negligible. As a matter of fact, the systematic absorption and transmittance are dominated by the population evolution in the $S_0$, $S_1$ and $T_1$ states, and the solution of Eq. (3) can be worked out with constant pumping condition as follows

$$n_{S0}(t) = n_0(0)[\gamma \exp(-\kappa t) - \kappa \exp(-\gamma t)]/(\kappa - \gamma)$$
$$n_{S1}(t) = n_0(0)\tau_{SG}[\exp(-\kappa t) - \exp(-\gamma t)]/(\kappa - \gamma) \qquad (11)$$
$$n_{T1}(t) = n_0(0) - n_{S0}(t) - n_{S1}(t),$$

where $\kappa$ and $\gamma$ are defined as

$$\kappa = \left[\eta - [\eta^2 - 4(\phi/\tau_{ISC})]^{1/2}\right]/2$$
$$\gamma = \left[\eta + [\eta^2 - 4(\phi/\tau_{ISC})]^{1/2}\right]/2$$
$$\eta = 1/\tau_{SG} + 1/\tau_{ISC} + (\sigma_G I/h\nu)$$
$$\phi = \sigma_G I/h\nu$$
(12)

To illustrate the time dependence of populations of mono-functional FCA and multi-functional Fol and MPF –containing polymers in the $S_0$, $S_1$ and $T_1$ states of the five-level model at various input irradiance operated with nano-second pulse duration, we simulate Eqs. (11) and (12) with the assumption that the nonlinear system is influenced either by strong excitation or by weak excitation. Figures 4(b) and 4(c) display for a relatively high intensity, the $S_0$ population is rapidly depleted, the $S_1$ population reaches a maximum in a short time and then decreases gradually, but the $T_1$ population builds up very quickly and finally reaches a steady value, which is expected to exceed the $S_0$ and $S_1$ populations. This simulation (figure 4(c)) is closely associated with the optical limiting processes of mono-and multi-functional fullerere/polymer composites observed in our experiments at high input fluences. Therefore, it is confirmed that the triplet-triplet state absorption can be effectively used for optical limiting besides singlet-singlet absorption. By adding more functional into $C_{60}$ cage as treated in multi-functional Fol and MPF, the π-conjugated electronic system becomes disturbed and this may cause a broken symmetry in the molecular system. Therefore, there would be a reduction in the lifetime of triplet-triplet state absorption as illustrated in Figure 4(b). However, it is evidently that by putting more polymers into the multi-functional, the slightly time enhancement of the triplet-triplet state was observed.

## 3.5. Polymer Concentration Dependence of Optical Nonlinearities in Supramolecular Fullerene

To keep the linear transmittance of the solutions constant, a series of cuvettes with different optical path lengths from 1 to 10 mm were used. At the highest concentration under consideration, ~$10^{-3}$ M, a cuvette with an optical path length of 1 mm was used. For solutions of lower concentrations, optical cells of longer path lengths were used in the measurements, for instance at concentration of ~$10^{-4}$ M with optical path lengths of 10 mm.

In figures 2 and 3, the polymers concentration dependence of supramolecular mono-functional FCA, and multi-functional FoI and MPF – containing polymers associated with optical limiting and photoluminescence responses were systematically measured and studied, respectively. As shown in Table 2, the optical limiting results are clearly dependent on polymer concentrations, with the changes particularly significant in the concentration range of 0.6 to 1.1 g/l for FCA/polymers, 0.5 to 1.0 mg/ml for FoI/polymers, and the ratio of 1:2 to 1:6 for MPF/PSI46, respectively. The highest polymer contributed to exhibit strong optical limiting responses, reaching a plateau at $F_{in}$ of ~0.7 J/cm$^2$ was observed in FCA/PSVPy32-A.

To check photostability of all the supramolecular samples, we have measured and compared the absorption spectra before and after laser irradiation. The obtained results indicated that there is no difference in the spectra for all the samples. Therefore, all the samples have a good photostability.

A closed correlation between either optical limiting and polymer concentration or optical limiting and the number of functional attached to $C_{60}$ cage in both the supramolecular mono-functional FCA and multi-functional Fol and MPF -polymers composites indicates that photo-excited electrons and their dynamic process play important role. This is different from the case of polymer/multi-walled carbon nanotube composites, in which polymers have negligible effects because of laser-induced nonlinear scattering - a completely different limiting mechanism [37,38].

**4. Conclusion**

In conclusion, the optical limiting, and PL measurements have been carried out to study the excited state mechanisms associated with the OL properties of supramolecular mono- and multi-functional $C_{60}$ - containing polymers. Moreover, the reverse saturable absorption in the five-level model that includes the prolongation of intersystem crossing and excited triplet state processes has been discussed for the OL performances of fullerene derivatives incorporated with polymers.

The FCA/polymer composites possess better OL behaviours than that of FCA, which are mainly contributed by stronger absorption of excited triplet state due to the presence of high concentration PSVPy32. Compared to $C_{60}$, Fol shows a weaker nonlinear optical response, which may be ascribed to the disturbance of π-electronic system of parent $C_{60}$ molecule due to multi-functionalization. The "inert" polystyrene does not significantly change the optical limiting performance of Fol solution, while the "active" PSVPy32 improves it. The MPF and MPF incorporated with polymer possess poorer nonlinear optical properties than that of $C_{60}$, which are mainly caused by the disturbance of π-electron system in $C_{60}$ cage due to multifunctionalization as well.

The OL and PL properties of supramolecular FCA, Fol and MPF – containing polymers are strongly dependent on the polymer concentration attributed to effects on the nonlinear optical mechanisms that are associated with excited triplet-triplet absorption processes. Further investigations using both time-resolved pump-probe and PL techniques to detect experimentally the time of both the first excited singlet-ground state and the triplet-triplet absorption processes, and to determine the related parameters are needed for proposing a quantitative modeling of the supramolecular fullerenes attached with polymers are needed.

**Acknowledgment**

We thank the National University of Singapore for the financial supports.

**Figure captions:**

**Figure 1.** UV-Vis absorption spectra of **(a)** 1: $C_{60}$ in toluene, 2: FCA, 3: FCA/PSVPy-A, and 4: FCA/PS-A, wherein, 2, 3 and 4 are in 1,2-dichlorobenzene solutions; **(b)** 5: toluene solutions of $C_{60}$-c, DMF solutions of 6: Fol-c, 7: Fol/PS-c, and 8: Fol/PSVPy32-$c_1$; **(c)** 9: $C_{60}$ in chlorobenzene, 10: MPF and 11: MPF/PSI-46(1:6) in tetrahydrofuran (THF), respectively, and 12: PSI-46. All the solutions are directly used in the optical limiting and Z-scan measurements at room temperature.

**Figure 2.** Photoluminescence spectra measured at room temperature of **(a)** 10-mm-thick $C_{60}$ in toluene, and FCA, FCA/PS-B, FCA/PSVPy32-B, FCA/PSVPy32-A and FCA/PS-A in 1,2-dichlorobenzene, respectively carried out at the same linear transmittance of 65% at 532 nm by using 440 nm as the excitation source; **(b)** 10-mm-thick Fol-d, Fol-c, **1**: Fol/PSVPy32-$d_1$, **2**: Fol/PSVPy32-$c_1$, **3**: Fol/PS-d, and **4**: Fol/PS-c in DMF, respectively; **(c)** 10-mm-thick $C_{60}$ in chlorobenzene, and MPF, MPF(1:2), MPF(1:4), and MPF(1:6) in tetrahydrofuran (THF), respectively. The PL spectra in **(b)** and **(c)** were conducted at the same excitation source of 480 nm.

**Figure 3.** Nonlinear transmission responses of **(a)** $C_{60}$-toluene (filled circles), FCA in 1.2-dichlorobenzene (open triangles), FCA/PSVPy32-B in 1.2-dichlorobenzene (filled diamonds), FCA/PSVPy32-A in 1.2-dichlorobenzene (open diamonds), and PSVPy32 in 1.2-dichlorobenzene (open inverted triangles); **(b)** toluene solutions of $C_{60}$-c (filled circles), and DMF solutions of Fol-c (filled squares), Fol/PSVPy32-$c_2$ (filled triangles),

Fol-d (open squares), and Fol/PSVPy32-d$_2$ (open triangles), respectively; **(c)** $C_{60}$ in chlorobenzene (filled circles), and MPF (open inverted triangles), MPF(1:2) (filled triangles), MPF(1:4) (open squares), MPF(1:6) (filled diamonds) and PSI-46 (open circles) in tetrahydrofuran (THF), respectively. The solid line is an example of the best fitting curve based on Eq. (8). The measurement of all the sample were conducted at the same wavelength of 532 nm.

**Figure 4.** **(a)** Schematic diagram of the five-level reverse saturable absorption model for dynamic processes of excited states in supramolecular mono- and multi-functional fullerene. The illustration of normalized population calculation as a function of time described the population dynamic (Eq. (3)) in supramolecular fullerene of the ground state ($n_{S0}$) and excited states ($n_{S1}$ and $n_{T1}$) with **(b)** weak excitation and **(c)** strong excitation, respectively.

**Figures and Tables**

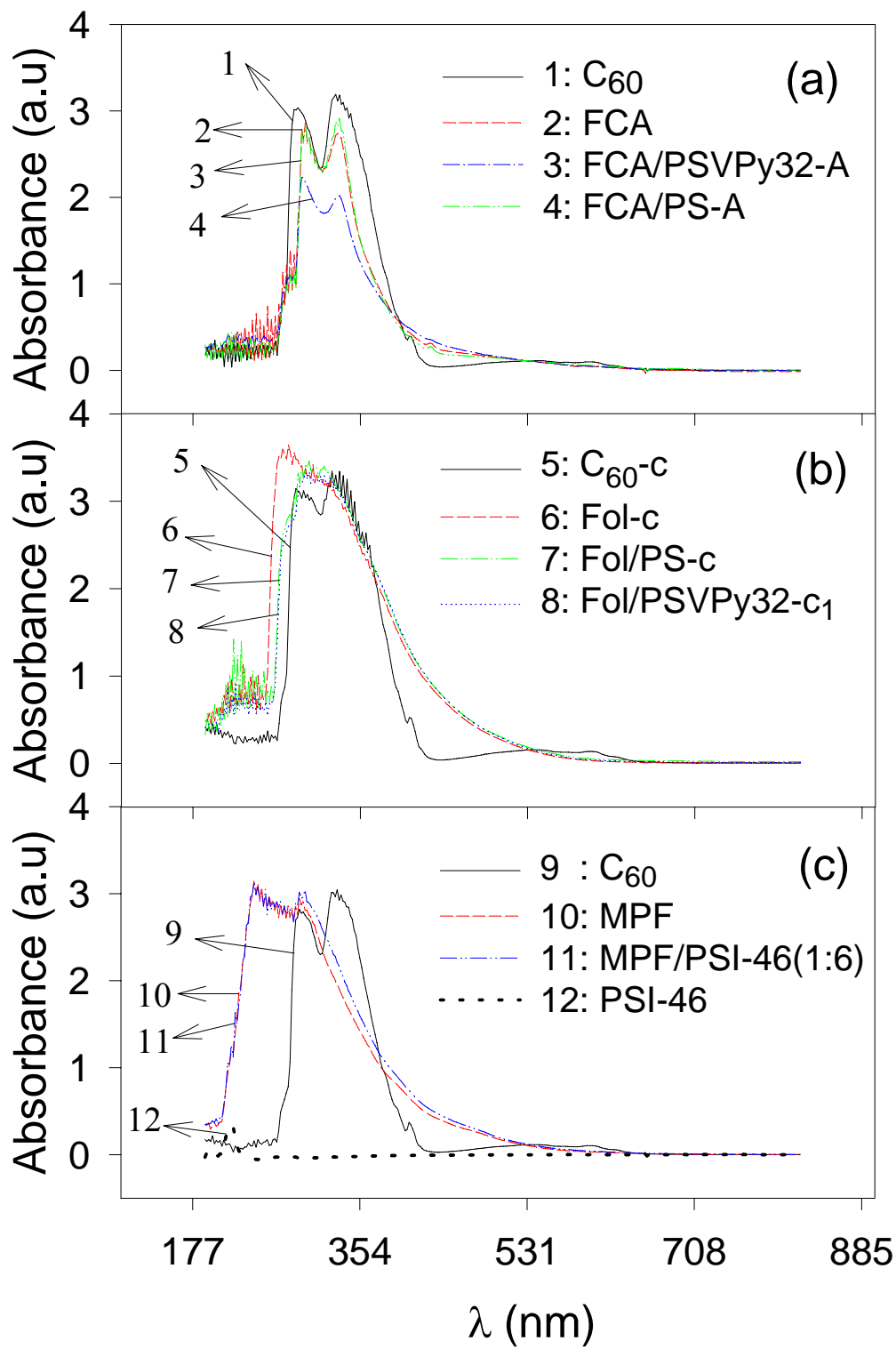

**Figure 1.**

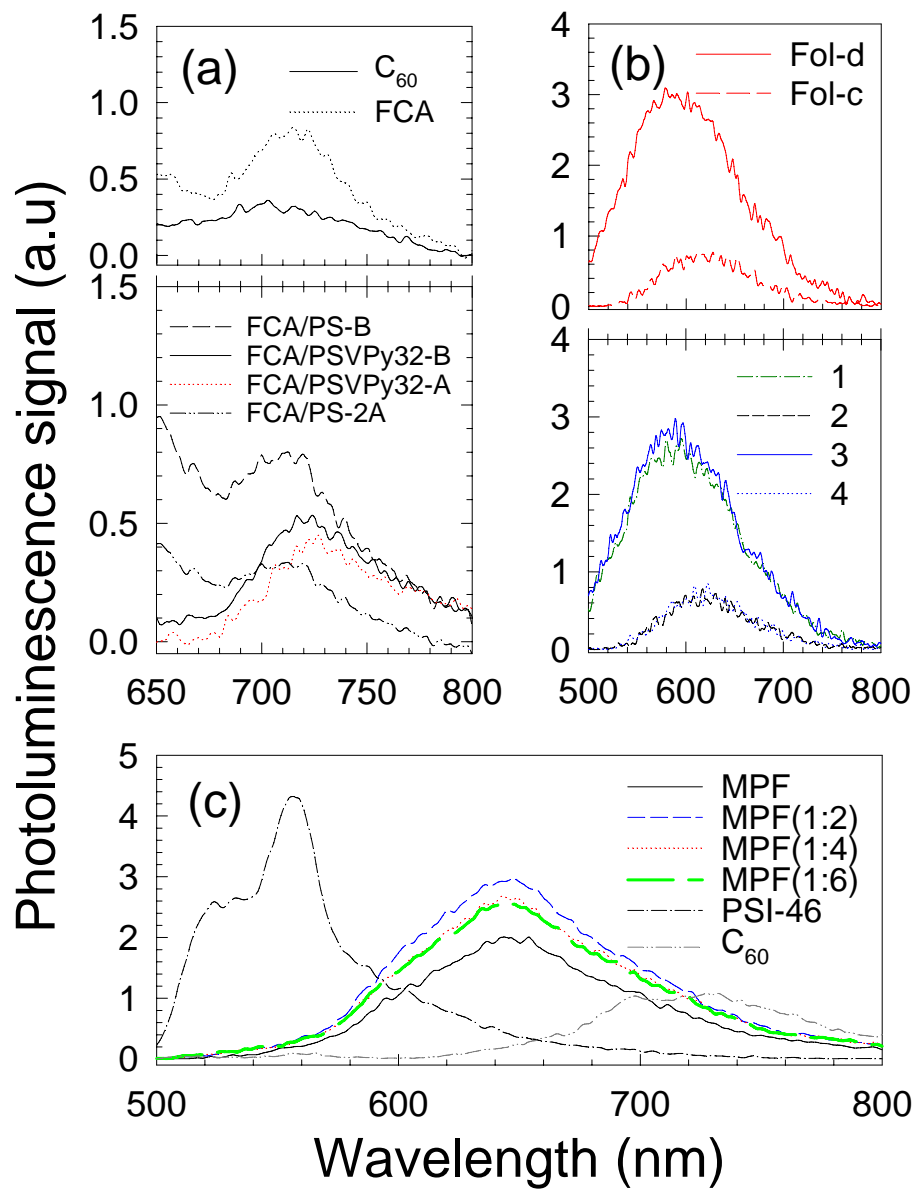

**Figure 2.**

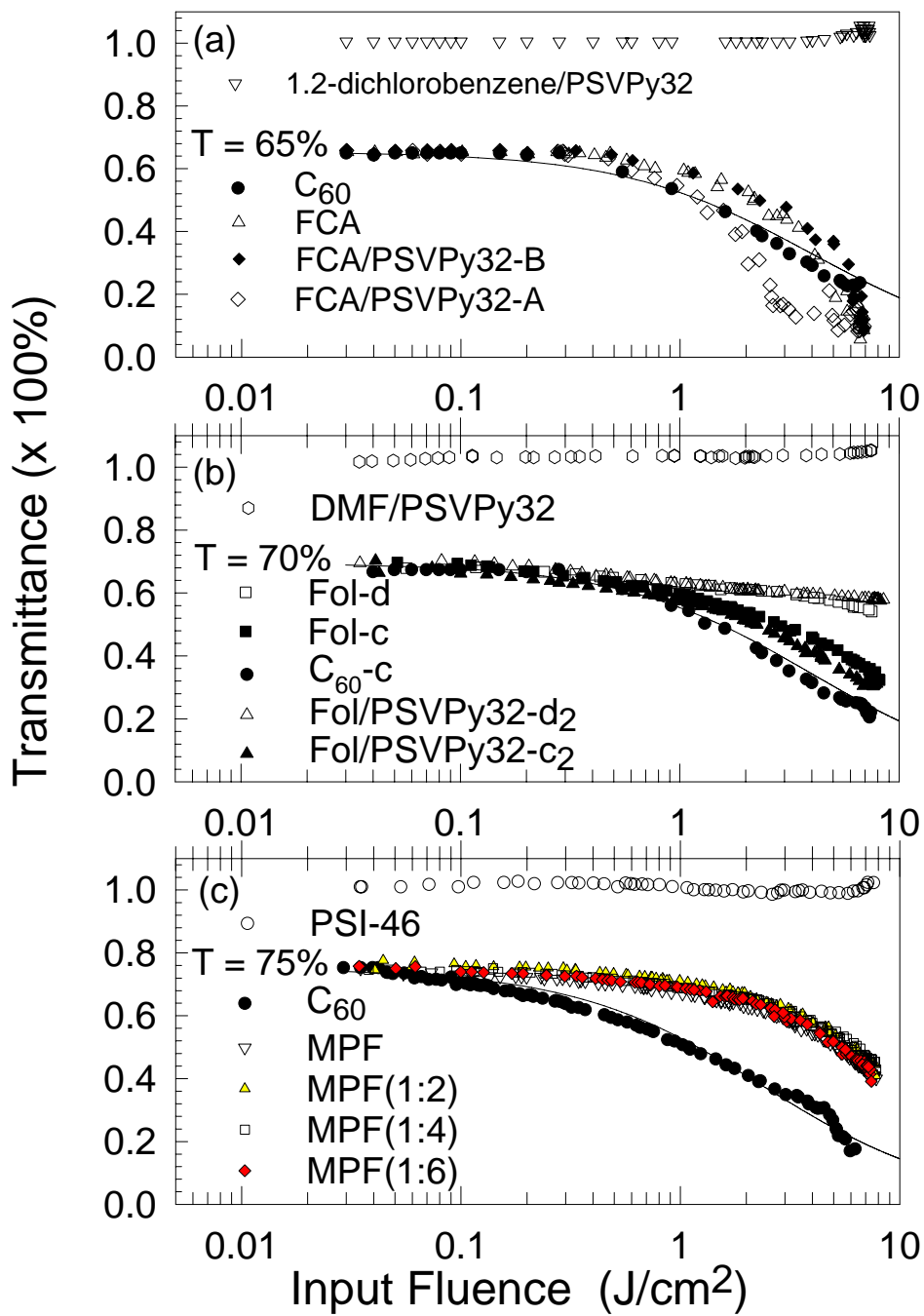

**Figure 3.**

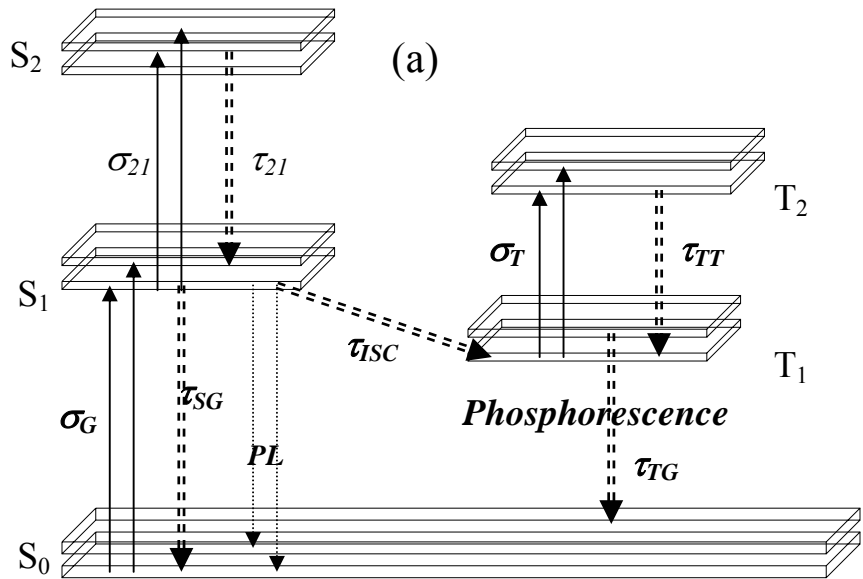
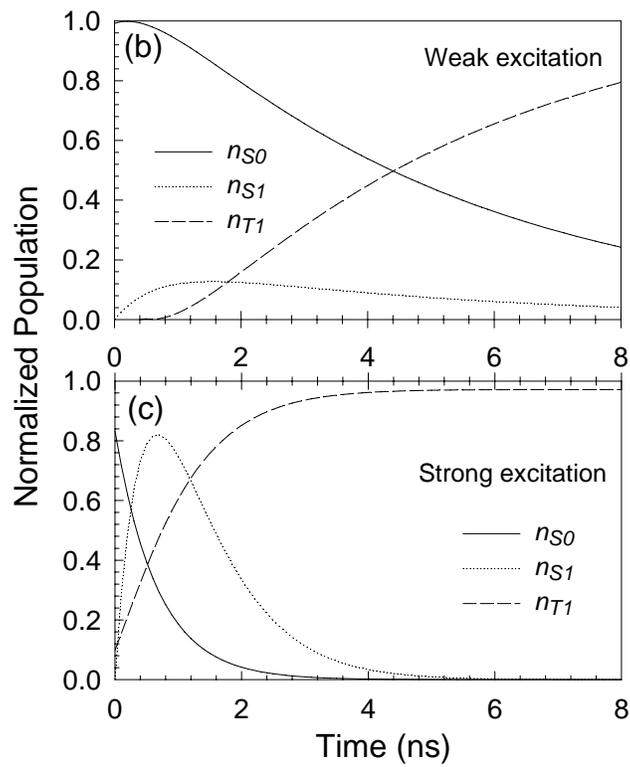

**Fig. 4.**

**Table 1.** Ground-state absorption parameters measured at 532 nm of $C_{60}$, and the supramolecular mono-functional FCA, and multi-functional Fol and MPF - containing polymers.

| Sample | Solvent | Concentration (M) | Linear Transmittance at 532 nm | Absorptivity $\varepsilon$ (M$^{-1}$ cm$^{-1}$) |
|---|---|---|---|---|
| $C_{60}$ | Toluene | 1.6 x10$^{-3}$ | 65% | 940 |
| FCA | 1.2-dichlorobenzene | 1.6 x10$^{-3}$ | 65% | 1169 |
| FCA/PSVPy32-A | 1.2-dichlorobenzene | 1.6 x10$^{-3}$ | 65% | 1169 |
| FCA/PSVPy32-B | 1.2-dichlorobenzene | 1.6 x10$^{-3}$ | 65% | 1169 |
| FCA/PS-A | 1.2-dichlorobenzene | 1.6 x10$^{-3}$ | 65% | 1169 |
| FCA/PS-B | 1.2-dichlorobenzene | 1.6 x10$^{-3}$ | 65% | 1169 |
| $C_{60}$-d | Toluene | 1.65x10$^{-4}$ | 70.1% | 940 |
| $C_{60}$-c | Toluene | 1.65 x10$^{-3}$ | 70.1% | 940 |
| Fol-d | DMF | 1.42 x10$^{-4}$ | 70.5% | 1069 |
| Fol/PSVPy32-d$_1$ | DMF | 1.42 x10$^{-4}$ | 69.6% | 1108 |
| Fol-c | DMF | 1.42 x10$^{-3}$ | 70.3% | 1080 |
| Fol/PSVPy32-c$_2$ | DMF | 1.42 x10$^{-3}$ | 69.5% | 1113 |
| $C_{60}$ | Chlorobenzene | 1.32 x 10$^{-3}$ | 75.3% | 940 |
| MPF | THF | 9.05 x 10$^{-3}$ | 75.2 | 140 |
| PSI-46 | THF | 14.0 gl$^{-1}$ | 100% | – |
| MPF(1:2) | THF | 9.05 x 10$^{-3}$ | 75.0% | 140 |
| MPF(1:4) | THF | 9.05 x 10$^{-3}$ | 74.7% | 140 |
| MPF(1:6) | THF | 9.05 x 10$^{-3}$ | 74.9% | 140 |

**Table 2.** Nonlinear five-level and limiting threshold parameters measured at 532 nm of $C_{60}$, and the supramolecular mono-functional FCA, and multi-functional Fol and MPF - containing polymers.

| Sample | Solvent | $\sigma_G$ (x $10^{-18}$ cm$^2$) | $F_{NLO}$ | $T_{SA}$ | $\sigma_{eff.}/\sigma_G$ | Limiting Threshold (J/cm$^2$) |
|---|---|---|---|---|---|---|
| $C_{60}$ | Toluene | 3.62 | 1.2 | 0.25 | 3.22 | 3.1 |
| FCA | 1.2-dichlorobenzene | 3.62 | 2.0 | 0.28 | 2.96 | 4.1 |
| FCA/PSVPy32-A | 1.2-dichlorobenzene | 3.62 | 0.7 | 0.15 | 4.40 | 2.3 |
| FCA/PSVPy32-B | 1.2-dichlorobenzene | 3.62 | 2.1 | 0.28 | 2.96 | 5.4 |
| FCA/PS-A | 1.2-dichlorobenzene | 3.62 | 0.7 | 0.15 | 4.40 | 2.0 |
| FCA/PS-B | 1.2-dichlorobenzene | 3.62 | 0.9 | 0.16 | 4.25 | 2.9 |
| $C_{60}$-d | Toluene | 2.77 | 1.6 | 0.32 | 3.19 | >5.2 |
| $C_{60}$-c | Toluene | 2.77 | 1.3 | 0.30 | 3.38 | 3.0 |
| Fol-d | DMF | 3.22 | 5.0 | >0.32 | <3.19 | >10 |
| Fol/PSVPy32-d$_1$ | DMF | 3.22 | 5.1 | >0.32 | <3.19 | >10 |
| Fol-c | DMF | 3.22 | 2.0 | >0.32 | <3.19 | >10 |
| Fol/PSVPy32-c$_2$ | DMF | 3.22 | 1.9 | >0.32 | <3.19 | >6 |
| $C_{60}$ | Chlorobenzene | 2.59 | 0.6 | 0.25 | 4.82 | 2.7 |
| MPF | THF | 0.38 | 2.0 | >0.4 | <3.18 | >10 |
| MPF(1:2) | THF | 0.38 | 2.4 | >0.4 | <3.18 | >10 |
| MPF(1:4) | THF | 0.38 | 2.3 | >0.4 | <3.18 | >10 |
| MPF(1:6) | THF | 0.38 | 2.2 | >0.4 | <3.18 | >10 |